\documentstyle[neuchatel]{article}
\begin {document}

%
%
%
{
\baselineskip 16pt
\begin{titlepage}

{\large\baselineskip18pt

\hfill CERN-TH/97-354  

\hfill KUL-TF-97/35

\hfill {\tt hep-th/9712135}

}

\begin{center}
\vfill
{\Large\bf D0-Branes as Instantons in D~=~4 Super Yang-Mills 
Theories~\footnotemark}
\footnotetext{\large ~To appear in the proceedings of the conference
{\sl ``Quantum Aspects of Gauge Theories, Supersymmetry and Unification''}, 
held at Neuch\^atel University, Neuch\^atel, Switzerland,
18--23  September 1997.}
\end{center}
\vskip .8in
\begin{center}
{\Large J. L. F. Barb\'on~$^\dagger$ and A. Pasquinucci~$^\ddagger$}

\vskip .4in

{\large\sl $^\dagger$ Theory Division, CERN

CH-1211 Geneva 23, Switzerland

{\tt barbon@mail.cern.ch} }

\vskip .14in

{\large\sl $^\ddagger$ Instituut voor theoretische fysica, K.U.\ Leuven

Celestijnenlaan 200D, B-3001 Leuven, Belgium}

\end{center}

\vskip1in

\begin{quotation}
\noindent {\Large\bf Abstract}:
{\large\rm We discuss the role of D0-branes
as instantons in the construction  of $SU(N)$ Super 
Yang-Mills and Super QCD
theories in four space-time dimensions 
with D4-, D6- and NS-branes. }

\end{quotation}
\vfill
\noindent {\large CERN-TH/97-354}

\noindent {\large December, 1997}
\vskip12pt
\end{titlepage}
}
\vfill\eject
\addtocounter{footnote}{-1}

%
%
%
\def\titleline{
D0-Branes as Instantons in 
D~=~4 Super Yang-Mills Theories 
}
\def\authors{
J. L. F. Barb\'on \1ad and A. Pasquinucci \2ad
}
\def\addresses{
\1ad
Theory Division, CERN\\
CH-1211 Geneva 23, Switzerland\\
\2ad
Instituut voor theoretische fysica, K.U.\ Leuven\\
Celestijnenlaan 200D, B-3001 Leuven, Belgium 
}
\def\abstracttext{
We discuss the role of D0-branes
as instantons in the construction  of $SU(N)$ Super 
Yang-Mills and Super QCD
theories in four space-time dimensions 
with D4-, D6- and NS-branes. 
}
\makefront


\def\npb#1#2#3{{\it Nucl. Phys.} {\bf B#1} (#2) #3 }
\def\plb#1#2#3{{\it Phys. Lett.} {\bf B#1} (#2) #3 }
\def\prd#1#2#3{{\it Phys. Rev. } {\bf D#1} (#2) #3 }

\def\bb#1{{\tt hep-th/#1}}

\def\CL{{\cal L}}   
   
\def\CM{{\cal M}}


\def\dj{\hbox{d\kern-0.347em \vrule width 0.3em height 1.252ex depth
-1.21ex \kern 0.051em}}

\def\half{{1\over 2}\,}

\def\Tr{{\rm Tr\,}}

\def\ket{\rangle}
\def\bra{\langle}

\def\tphi{\widetilde \phi}
\def\tH{\widetilde H}
\def\tpsi{\widetilde \psi}
\def\tZ{\widetilde Z}
\def\tz{\tilde z}

\def\semi{; }

\newcommand{\beq}{\begin{equation}}
\newcommand{\eeq}{\end{equation}}
\newcommand{\ben}{\begin{equation*}}
\newcommand{\een}{\end{equation*}}
\newcommand{\baq}{\begin{eqnarray}}
\newcommand{\eaq}{\end{eqnarray}}
\newcommand{\ban}{\begin{eqnarray*}}
\newcommand{\ean}{\end{eqnarray*}}
\newcommand{\brr}{\begin{array}}
\newcommand{\err}{\end{array}}
\newcommand{\bc}{\begin{center}}
\newcommand{\ec}{\end{center}}

\newcommand{\bea}{\begin{eqnarray}}
\newcommand{\eea}{\end{eqnarray}}
\newcommand{\bean}{\begin{eqnarray*}}
\newcommand{\eean}{\end{eqnarray*}}


%
%

The basic construction of SYM instantons on a D$p$-brane world-volume, 
is as bound states with D$(p-4)$-branes \cite{rwinst}.
In the case of the construction of refs.\ \cite{regk}
(see also \cite{rberk}), 
the $SU(n_c)$ SYM theory
is described by $n_c$ D4-branes, so we must consider D0-branes
bound to D4-branes. The  five-dimensional
world-volume of the D4-branes is then compactified 
to the four-dimensional physical space and, to make a bound D0-brane 
describe an instanton in the four-dimensional physical space, the 
D0-brane Euclidean-time world-line must lie along the compactified 
direction of the D4-brane.  Although the
details of such a reduction  
depend on the amount of supersymmetry  ($N=4,2,1,0$) that we want
to preserve in space-time, the basic properties of the
instanton moduli space only depend on the geometry of D0-D4-branes
bound states, which we now review.      

The collective dynamics of the instantons is specified by a point
action and an appropriate measure over instanton moduli space. In
the present D0-branes construction, both ingredients appear as the
dimensional reduction down to zero dimensions of the D0-brane 
one-dimensional collective dynamics. In particular, the instanton 
moduli space as a hyper-K\"ahler quotient appears as a
Higgs branch of the world-line theory of the D0-branes.
The basic terms in the world-line action are 
the static Born--Infeld action of the D0-brane
\beq
S_{BI} = {\tau_0 } \int \sqrt{{\rm det}\, h_{\rm induced}}\ \longrightarrow\
M_{D0} \int d\tau,
\label{bi}
\eeq
and a source term coming from the Chern-Simons couplings on
the D4-branes world-volume 
\beq
S_{\rm source} = i {\mu_4 \over 2}
 \int_{\Sigma_{4+1}} A_{D0}^{RR} \,(2\pi\alpha')^2 \,
F\wedge F,
\label{source}
\eeq  
which relates the RR photon of the type-IIA string theory
to the theta parameter of the four-dimensional
instanton.
The basic D0-brane action (\ref{bi}) leads to a fit of the effective
length of the world-line, in order to reproduce the instanton
action $S_{\rm inst} = 8\pi^2 /g^2$. Indeed from the Born--Infeld 
action of the D4-brane gauge theory, upon dimensional reduction
in the world-line direction (say, the $x^6$ direction), on an interval
of length $L_6$,  we have 
\beq
{\tau_4 } \int_{0}^{L_6} d\tau \,\Tr\,\sqrt{{\rm det}(
1+2\pi\alpha' F)} \rightarrow -{\tau_4 \over 4} L_6 (2\pi\alpha')^2
 \,\Tr\,F^2 \equiv
-{1\over 4g^2_{YM}} \, \Tr\, F^2.
\label{dimrdf}
\eeq
Using $\tau_p^{-1} = g_s \sqrt{\alpha'} (2\pi \sqrt{\alpha'})^p$ and
$M_{D0} = \tau_0$ we find agreement since 
$8\pi^2/g^2\equiv M_{D0} L_6 =8\pi^2/g^2_{YM}$.
Also, for a constant one-form $A_{D0}^{RR}$, the five-dimensional 
integral (\ref{source})\ is localized 
on the D0-brane world-line, and   we have 
$S_{\rm source} =-i {\theta \over L_6} \int d\tau $.

The zero-mode structure of the instantons comes
 from the non-trivial world-volume dynamics
of the D0-branes,  as given by the massless limit of the
corresponding open strings.  
There are non-trivial gauge interactions specifying the dynamics of 
0-0 strings and 0-4 strings, in the general case where we have $k$ D0-branes 
bound to $n_c$ D4-branes.  
The 0-0 sector is just the dimensional reduction down to one
dimension of ten-dimensional SYM with $U(k)$ gauge group, corresponding to
a system of $k$ D0-branes, and has $N=4$ supersymmetry in four-dimensional
 notation. The
0-4 sector breaks half of these unbroken 16 supercharges, leading
to an action for the 0-4 strings with $N=2$ supersymmetric couplings.

To be more explicit, let us parametrize the different degrees of
freedom in four-dimensional superfield notation, i.e.\ we define
the world-line action by dimensional reduction from an
$N=2$ model in four dimensions, which can be obtained as a T-dual 
configuration of D3-branes inside D7-branes of type IIB string theory.
This auxiliary four-dimensional space is along the $(x^6, x^7, x^8, x^9)$
directions. 
We denote by $M$ the physical Minkowski space-time $(x^0, x^1, x^2, x^3)$
and after a Wick rotation, 
we shall work with complex coordinates $z=x^0 +ix^1$, 
${\tilde z} =x^2 +ix^3$  and $w=x^4 +ix^5$.
The 0-0 matrix coordinates, representing generalized translational
degrees of freedom of the $k$ D0-branes, are described by a set of chiral
superfields $\Phi_{00} \equiv (W, Z, {\widetilde Z})$, representing
displacements in the corresponding coordinates, and by the dimensionally
reduced vector superfield $V_{00} \sim (A_6, A_7, A_8, A_9)$, plus
superpartners, all in the adjoint representation of $U(k)$.  
We will not consider
excitations or deformations of the field $A_7$, and we will gauge
the longitudinal $A_6$ to zero (recall that the D0-branes world-lines
lie in the $x^6$ direction). Finally, the 0-4 and 4-0 sectors give rise
to hypermultiplets, represented by chiral superfields 
$\Phi_{04} =H, \Phi_{40}= \tH$,
transforming in the $(n_k, {\bar n}_c), ({\bar n}_k, n_c)$ of 
$U(k)\times U(n_c)$ respectively. The complete superpotential contains
an $N=4$ interaction between 0-0 fields:   
\beq
{\cal W}_{0-0} = {\sqrt{2} \over  g_s}  \, {1\over 3!}\,
\epsilon_{ijk}
\,\Tr\, \Phi_i [\Phi_j, \Phi_k]
\label{oosup}
\eeq
and an $N=2$ preserving superpotential for the ``quarks": 
\beq
{{\cal W}_{0-4} = {\sqrt{2}\over g_s}  \, ( \tH W H -
H W' \tH ), }
\label{ocsup}
\eeq
where we have introduced a mass term inherited from the superpotential
in the D4-branes, i.e.\ the 4-4 strings, coupling the $SU(n_c)$ adjoint
$W'_{44} = X'_4 + iX'_5$ giving the position of D4-branes. The $N=2$ 
supersymmetry
on the D0-brane world-line implies that $[W', W'^{\dagger}]=0$, i.e.\ the
$W'$ field lies along the flat directions of the D4-brane theory, and can
be taken as diagonal matrices by an $SU(n_c)$ rotation\footnotemark
\footnotetext{The brane
construction of the full gauge theory produces naively a $U(n_c)$ group.
Only the $SU(n_c)$ subgroup is relevant to the instanton discussion,
of course. In fact, on the Coulomb branch of the D4-branes world-volume,
the extra $U(1)$ is frozen, since fluctuations of the trace of $W'$
have infinite norm \cite{rwittenM}.}   
$W'^b_a = w'_a \,\delta^b_a$. This lagrangian also appears in various 
M-atrix theory applications \cite{rkachru}.

The instanton moduli space is then obtained as a maximal Higgs phase  
of this theory on the D0-brane world-line, where we consider the most
general vacuum expectation values of the hypermultiplets $H, \tH$ and
the adjoint fields $Z, \tZ$. In this paper we will be mainly concerned
with properties of the single instanton sector, or equivalently, of the
dilute gas limit of instantons. 
In this case the  overlap  between the instantons
 can be neglected and the moduli space admits a much simpler
description.  Indeed, well separated D0-branes are  
characterized by the complete diagonalization of all the
adjoint 0-0 fields corresponding to the space-time positions, i.e.\ we
consider $Z, \tZ$ diagonal generic matrices, which spontaneously break
$U(k)$ to the Abelian subgroup $U(1)^k$. It is intuitively clear that
this produces just $k$ copies of the single instanton moduli space.   
In this subspace, the $4k$ position degrees of freedom $z_i , \tz_i$ are
completely decoupled from the rest of the variables.  

The interesting branch of solutions is the one with maximal Higgsing of
the hypermultiplets, with ${\rm diag}(W)={\rm diag} (W') =0$ 
(all D0 and D4-branes on top of each other in the $(x^4, x^5)$ plane).  
Subtracting the $2k$ $F$-equations
and the $k$ $D$-equations from the total $4kn_c$ real components in
the scalars $\phi_h, \tphi_h$, and further dividing by the $U(1)^k$
gauge symmetry, we find a total of $4kn_c -2k-k-k = 4k(n_c -1)$ real
parameters. Adding the $4k$ degrees of freedom coming from the eigenvalues
of $Z$ and $\tZ$, representing the translation moduli in the physical
dimensions, we end up with the correct $4kn_c$ dimensional bosonic moduli 
space of YM instantons.

%
%
\vskip6pt

The basic idea in this microscopic model for YM instantons is
then the Kaluza--Klein truncation of the D4 gauge theory with
16 supercharges, down to a four-dimensional theory, 
together with a zero-mode truncation of 
the world-line dynamics of the D0-branes.  

In the simplest case, upon reduction on a circle of length $L_6$,
we get $N=4$ SYM in four dimensions. On the relevant moduli space
of the D0-branes, we are instructed to keep only the zero modes.
For $k$ D0-branes,  the 
 world-line has a set  of $d=2kn_c$ free superfields $(\xi_s, \eta_s)$, 
in four-dimensional notation, i.e.\ we have an action
\beq
S_{D0} =k M_{D0} \int_0^{L_6} d\tau - ik\theta + {1\over g_s} 
\int_0^{L_6} d\tau
\sum_{s=1}^{d} \left( |{\dot \xi}_s |^2 + i
 \,{\overline \eta}_s {\dot \eta}_s 
\right) + {\rm massive\ .} 
\label{accl}
\eeq
Each  Weyl fermion $\eta_s$ has two complex or 
four real components off-shell so that, when the
one-instanton contribution to the path-integral is dominated by the
classical static trajectory ${\dot\xi}_s = {\dot \eta}_s =0$,  we
get a factor
\beq
Z_{\rm inst} = \int \prod_s d{\overline\eta}_s d\eta_s \, d\mu(\xi_s)\,
e^{-kM_{D0} L_6 + ik \theta
} = 0^{4d} \, {\rm Vol}({\CM}) \, e^{-{8\pi^2 k \over g^2} 
+ik\theta}
\label{instf}
\eeq
where $M_{D0} L_6 \equiv 8\pi^2 / g^2$. 
In this expression, the bosonic measure $d\mu(\xi)$
for collective coordinates contains zero-mode Jacobians  which are most
easily derived in the field theoretical framework. The corresponding
integral gives the formally divergent volume of moduli space, 
${\rm Vol} (\CM )$. The symbol $0^{4d}$
denotes the number of fermionic zero modes of the instanton, to be 
saturated by fermion sources. We see clearly that the number of zero
modes corresponds to the total number of independent anticommuting 
collective coordinates, which in turn coincide with the total number
of off-shell fermion components on the D0-brane world-line.
The corresponding number of fermion zero-modes is $4d=8kn_c$, 
 which is indeed 
the right one for $SU(n_c)$ SYM in the $k$-instanton sector. In field theory, 
each gluino system gives $2kn_c$ zero modes, and we have four independent
gluino systems in $N=4$ SYM. In the D0-D4 branes construction,
the $8k$ zero modes associated to the
$Z, \tZ$ superfields are interpreted in the four-dimensional SYM theory
as supersymmetry zero modes. On the other hand, $8k$ out of the $8k(n_c-1)$
zero modes associated
to the hypermultiplets $H, \tH$ are interpreted in space time as
superconformal zero modes, while the remaining  $8k(n_c -2)$ zero modes
do not have an obvious symmetry interpretation, since they arise
as 't Hooft zero modes associated to doublets with respect to the
$SU(2)$ subgroup of the gauge group where the instanton sits.

According to the constructions of refs.\ \cite{regk}, 
models with $N=2$ and $N=1$ supersymmetry can be obtained upon compactification
on a segment of length $L_6$, bounded by two NS5-branes. One of them, denoted 
NS, extends in $(M, w)$, and the other, NS$^\prime$, possibly rotated, 
extends in $(M, \Pi)$, with $\Pi$ a plane in the 
$(w,v)$ space. In the case of $N=2$ supersymmetry the NS and NS$^\prime$ 
branes are parallel, whereas for $N=1$ the plane $\Pi$ is
obtained from the $w$-plane by an $SU(2)$ rotation.  
For the purposes of the D0-branes dynamics, the role of the NS5-branes is
just to project out some fermionic zero modes, namely, those zero modes
among the total of $8kn_c$ anticommuting coordinates $\eta_s, {\overline
\eta}_s$ that overlap with broken supersymmetries of the brane configuration
in $M$. These zero modes correspond really to Goldstone fermions and are not
normalizable in the supersymmetric vacuum of $M$. Further, we assume
that no projection is imposed by the NS5-branes on the bosonic collective
coordinates of the D0-branes. Indeed to the extent that they are bound to the
D4-branes, they can freely move in (the Euclidean version of) $M$, and 
furthermore the Higgs branch collective coordinates do not feel 
any geometric constraint as a result of the presence of the NS5-branes.

In order to see how the fermion projections work, we recall that the
unbroken supersymmetry on a brane world-volume is best represented in
M-theory language in terms of the projectors  
$P_p = {1\over 2} (1+i \Gamma (\Sigma_{p+1}))$, 
where $\Gamma (\Sigma_{p+1})$ is the product of Dirac matrices in the
world-volume directions, with the understanding that $p$-branes whose
M-theory description involves a wrapped or boosted M-brane in the
eleventh direction include a factor of $\Gamma_{11}$. 
So, in terms of a 32-dimensional
spinor $\Psi_{32}$, the 16 Goldstone fermions carried by a free D0-brane
are given by $(1-P_{D0})\Psi_{32}$.
Now, according to  the above
prescription, we should keep only those zero modes that have no overlap
with the Goldstone fermions on the NS-brane world-volume. The remaining zero
modes from the 0-0 sector of a general free D0-brane  are then 
$ \eta_{z, {\tilde z}, v, w}
 = (1-P_{D0}) P_{NS} P_{NS'} \Psi_{32}$. 
Using the explicit form of the projectors, one easily sees that each
NS-brane divides by half the number of zero modes, so that we have 8
zero modes in the $N=2$ configuration and 4 zero modes in the $N=1$
configuration.

 The previous considerations apply to the situation of a free D0-brane.   
In the case of a D0-D4-branes bound state we have two modifications. First,
as stated before, on the Higgs branch of bound states the fermions
$\psi_v , \psi_w$ are lifted, and the number of zero modes coming
from the 0-0 sector is again divided by half. Second, we have a new
set  of degrees of freedom, namely the $\psi_h, \tpsi_h$ from the
0-4 and 4-0 sectors. Those are constructed   
 from the zero modes of the world-sheet fermions $\psi_0^{\mu}$ for 
$\mu=4,5,6,7,8,9$ (i.e.\ in the NN $+$ DD directions of the D0-D4-brane
system), in the Ramond sector.  
Indeed, there are $\half 2^{6/2} = 4$ field components (off shell), after
GSO projection, in the 0-4 sector, that is a Weyl fermion. In terms of the
complex combinations $\Gamma_u = \half (\Gamma_6 +i\Gamma_7)$, $
\Gamma_w = \half (\Gamma_4 +i\Gamma_5)$, $\Gamma_v = \half (\Gamma_8 +
i\Gamma_9)$, the chirality operator reads
\beq
\Gamma_{11} = i\Gamma_0 \Gamma_1 \cdots \Gamma_9 =-(1-2N_z)(1-2N_{\tilde z})
(1-2N_u)(1-2N_v)(1-2N_w), 
\label{gsop}
\eeq
with $N_v \equiv \Gamma_v^{\dagger} \Gamma_v $, etc., the fermion occupation
numbers taking values 0 or 1. Acting on the $2^{6/2} + 2^{6/2}$ states in
0-4 and 4-0 sectors, built as polynomials $P(\Gamma_u^{\dagger},
\Gamma_v^{\dagger}, \Gamma_w^{\dagger} ) |0\ket$, the GSO projection
reduces to $-(1-2N_u)(1-2N_v)(1-2N_w) =+1$. On the same set, the 
projection imposed by the NS brane is $+1 = i\Gamma_{NS} = i\Gamma_0 
\Gamma_1 \cdots \Gamma_4 \Gamma_5 = -(1-2N_w)$. So, on this subspace
the NS projector is $P_{NS} = N_w$, and the analogous
projector for the rotated NS$^\prime$-brane is $P_{NS'} = N_v$.  

We see that, in all cases, the effect of each NS-brane 
is to divide by 2 all the bulk zero modes democratically.  
This amounts to a total factor of $1/4$ when the NS$^\prime$ 
brane is a rotation
of the NS-brane into the $(x^8, x^9)$ plane. So,
if $N$ denotes the amount of space-time
supersymmetry of the brane configuration, the resulting instanton
leaves $2N$ supercharges unbroken, out of the total $4N$ that are 
linearly realized in the vacuum of the space-time theory. 
The total number of fermionic zero-modes is given by:
\beq
{\cal N}_{f.z.m.} = {N\over 4} \cdot 4d 
  = {N\over 4}\cdot  8 k + {N\over 4}\cdot 8k(n_c-1) = 2Nkn_c,  
\label{totz}
\eeq
the correct number, where
we have already subtracted the massive degrees of freedom in the
mixed Higgs-Coulomb branch, which defines the instanton moduli space,
and we have split the zero modes between those corresponding to
the $2N$ broken supercharges,  and those of a fixed instanton, coming from
the hypermultiplet degrees of freedom.  

Introducing flavour brings in new subtleties. Now we have $n_f$ D6-branes
extended in $(M, v,  x^7)$, and localized at definite values
of $x^6$. Since the D0-branes world-line extends in $x^6$, the local
D0-D6 intersections are instantonic! This means that the 0-6 and
6-0 strings must be quantized with boundary conditions such that
$\nu = d_{ND} + d_{DN} =8$ (the sum of ND and DN directions). Thus the
0-6 sector is T-dual to a D1-D9 sector, and massless fermionic
degrees of freedom appear localized at the D0-D6 intersections.   
Carrying out the analysis \cite{rus}, one finds the required 
$2kn_f$ fermionic
zero modes associated to the quarks in the space-time description. Since
these zero modes are localized at intermediate points in the D0
world-line, they are insensitive to the projections imposed by the
NS5-branes at the end-points. Therefore, the number of flavour zero
modes depends only on $n_f$, but not on the amount of space-time
supersymmetry, in beautiful correspondence with the field theory result. 
%
%
%
\vskip6pt

Up to now we have considered the case of exact instantons in super Yang-Mills 
theory and showed that it can be described by a system of $k$ D0-branes 
bound to $n_c$ D4-branes. In particular this means that the D0-branes and
the D4-branes are all located at the same point in the 
$(x^4,x^5,x^7,x^8,x^9)$ plane. The
geometrical operation of moving the D4-branes apart corresponds, as 
intuition would tell, to describing constrained instantons
\cite{rtof}.
%
%
The relevant coupling in the D0-brane world-line theory,
which is sensitive to  the D4-branes motion is (see eq.\ (\ref{ocsup}))
\beq
\CL_{04-44-40} = {\sqrt{2}\over g_s}
\int d^2\theta \,\Phi_{04} \Phi_{44} \Phi_{40}
+ {\rm h.c.}
\label{lfour}
\eeq
This coupling is simply a mass term (the mass matrix given by
$\phi_{44} \sim W'$ that describes the position of the D4-branes in the 
$(x^4,x^5)$ plane) for the instanton moduli (in the notation of
(\ref{ocsup})\ 
$\Phi_{04} \sim \tH$ and $\Phi_{40} \sim H$). When considered in the
Coulomb branch of the $N=2$ configurations, it yields the familiar
action of a constrained instanton:
\beq
S_{\rm constr} = {1\over g^2} (8\pi^2 + \rho^2 v^2 ) \ ,
\label{Sconstr}
\eeq
with $v \sim \bra \phi_{44} \ket$ and $\rho^2 \sim \phi_{04}\phi_{40}$.
At the same time, it lifts the fermion zero modes through the
pairings $v \psi_{04}\psi_{40}$. For the fermions, the lifting of
zero modes occurs homogeneously in the bulk of the D0-brane world-line, 
and therefore it is obviously compatible with the projections imposed 
by the NS-branes at the boundaries. 
In general, a homogeneous lifting pattern of this sort, appropriate
to describe the Coulomb phase of $N=2$ models, gives a mass of order
$v$ to $d_c$ complex 
bosonic collective coordinates, and similarly, it lifts
$4d_c$  real fermionic coordinates in the bulk, i.e.\ it induces
 a term in
the bulk action of the form
$
\delta\CL \sim v^2 \sum_{s=1}^{d_c} |\xi_s |^2 + v\sum_{s=1}^{d_c} 
{\overline \eta}_s \eta_s$,  
where we have rescaled the bare Yang-Mills coupling $g^2 \sim g_s /L_6$ 
into the collective coordinates $\xi_s$ and $\eta_s$.   
Taking into account the projection at the boundaries of the world-line 
due to the NS- and NS$^\prime$-branes, we have a partition function of the form
\beq
Z_{\rm inst}
 \sim 0^{N(d-d_c)}\,\, {\rm Vol}_c (\CM) \,\,v^{Nd_c \over 2}\,\,  
e^{-{8\pi^2 \over g^2} +i \theta } 
\label{Zconstr}
\eeq
where ${\rm Vol}_c (\CM)$ represents the bosonic ``volume" of the
moduli space, regulated with the Gaussian term ${\rm exp}(-v^2 |\xi|^2)$,
and the term $v^{Nd_c /2} $ comes from the fermionic integrals.\footnotemark
\footnotetext{
Recall that $N$ is the number of four-dimensional
supersymmetries in the space-time
configuration, and $d =2n_c$ is the complex dimension of the 
bosonic instanton moduli space.} 
The complete power of $v$ in the full instanton
measure depends on the result of the bosonic integrals, of course.
In general, we must view the couplings like eq.\ (\ref{lfour})\  as small
perturbations over the instanton moduli space described in the 
introduction. This amounts to the requirement that the dimensionless
effective expansion parameter $\rho v$ be small.

The important remark to make is that all zero modes due to the 
conformal symmetry and embedding angles, namely, all zero modes
encoded in $\Phi_{04}, \Phi_{40}$, can be lifted in this way on the
Coulomb branch of the $N=2$ theory, by switching on $\Phi_{44}$ (i.e.\      
splitting the D4-branes in the $(x^4, x^5)$ plane). This amounts
to generically  lifting
$2d_c = 4(n_c -1)$ bosonic and $4(n_c -1)$ fermionic zero modes.
The remaining zero modes, $4$ bosonic and $4$ fermionic, correspond
to the superfields from the 0-0 sector $Z$ and $\tZ$, which remain
uncoupled and therefore are not lifted. Since these collective 
coordinates are
associated to the translations in the physical four dimensions,
these fermion zero modes are naturally interpreted as
due to the broken supersymmetries of the instanton, understood
as a classical solution in the physical four-dimensional
theory on the four-branes.
 
Using these ideas, one can postulate simple couplings at the
intersections between D0-, D4- and D6-branes, which explain the
pattern of zero-mode lifting appropriate to $N=2$ and $N=1$ 
Higgs branches. Analogous, even if not so detailed, considerations
can be done also in the case of multi-instanton configurations beyond 
the dilute-gas approximations. We refer the reader to \cite{rus} for 
more details.

%
%
\vskip6pt

Although our discussion applies strictly
 to the regime of weak type-IIA coupling,
it is interesting to ask ourselves
what happens of the D0-branes we have been discussing in this paper, in
the lifting to M-theory \cite{rwittenM}.
A free D0-brane of the type-IIA theory  
becomes a graviton Kaluza--Klein                 
 state carrying momentum along the circle in the eleventh direction,  
and  electrically charged with respect to the $U(1)$ gauge field 
corresponding to the rotations around this circle. As discussed in
 refs.\ \cite{rwitqcd},
the physics of such 
free D0-branes seems to have nothing to do with the field theory
SYM models we are interested in. Whereas in the M-theory picture it could be
less simple to decouple these modes, in the type-IIA construction we can
just declare that we do not consider free D0-branes. In this way, we are
making an explicit distinction between the free D0-branes, and the
D0-D4 branes bound states. Indeed this distinction is 
natural from the point of view of the type-IIA configuration, since 
there is no non-contractible circle where we could wrap a free D0-brane
world-line to 
yield a semiclassical instanton contribution. On the other hand, the
world-line of the bound D0-brane is just the Euclidean world-line
of the D4-brane with which it is solidary. In other words, once the
D0-branes are bound to the D4-branes, they must be considered as part
of the dynamical data on the D4-branes world-volume.                

Thus, the  only D0-branes
of interest are those that are bound to at least two D4-branes and are not
allowed to become free. In the internal space, the D0-branes just completely
adhere to the D4-branes since they differ only in the space-time directions
where the D0-branes are at a point. In lifting to M-theory, only the 
internal space gets modified and from this point of view the bound 
D0-branes just completely merge in the D4-branes and they become part
of the resulting five-brane. Actually one could see this from the 
opposite point of view: in extending the D4-branes to become an M-theory
five-brane wrapped around the eleventh circle, 
one is effectively dressing them up with bound D0-branes.   
Further discussion of theta angle dependence in MQCD can be found in ref.\
\cite{rwe}.

%
%
\vskip18pt
\noindent{\Large\bf Acknowledgements}
\vskip6pt

\noindent
This work is partially supported by the European Commission TMR programme
ERBFMRX-CT96-0045 in which A.P.\ is associated to the Institute for 
Theoretical Physics, K.U.\ Leuven. A.P. would like to thank CERN for its
hospitality while part of this work was carried out.


\end{document}